\documentclass[12pt]{article}

\usepackage{amssymb,latexsym,amsmath} 
\usepackage{authblk}
\usepackage[utf8]{inputenc}
\usepackage{dsfont}
\usepackage[utf8]{inputenc}
\usepackage{graphics}
\usepackage{color}
\usepackage{subfig}
\usepackage{graphicx}
\usepackage[colorlinks = true,
linkcolor = blue,
urlcolor  = blue,
citecolor = blue,
anchorcolor = blue]{hyperref}
\usepackage[numbers,square,sort&compress]{natbib}

\begin{document}
\title{Circuit Complexity in $U(1)$ Gauge Theory}

\author{Amir Moghimnejad \thanks{a.moghimnejad@modares.ac.ir} }
\author{Shahrokh Parvizi \thanks{parvizi@modares.ac.ir} }

\affil{Department of Physics, School of Sciences,
	Tarbiat Modares University, P.O.Box 14155-4838, Tehran, Iran}

\maketitle
\begin{abstract} 
	We study circuit complexity for a free vector field of a $U(1)$ gauge theory in Coulomb gauge, and Gaussian states. We introduce a quantum circuit model with Gaussian states, including reference and target states. 
	Using the Nielsen's geometric approach, the complexity then can be found as the shortest geodesic in the space of states. This geodesic is based on the notion of geodesic distance on the Lie group of Bogoliubov transformations equipped with a right-invariant metric. We use the framework of the covariance matrix to compute circuit complexity between Gaussian states. We apply this framework to the free vector field in general dimensions where we compute the circuit complexity of the ground state of the Hamiltonian.
\end{abstract}

\section{Introduction}\label{sec:intro}
In the recent years, the concept of complexity in physical systems has been inspired by the computer science and conceptually is how difficult to reach a target state starting from a certain reference state. This of course needs to be defined more rigorously which comes in the following. The importance of complexity is in the AdS/CFT correspondence where one can study the emergence of spacetime from field theory degrees of freedom \cite{ref1, ref2, ref3, ref4}. It is also related to the entanglement entropy \cite{ref5}. In this context, the complexity also provides a new tools to probe the interior of a black hole. There are two well-known conjectures as holographic duals to the complexity. The first one is the complexity=volume (CV) in which the complexity is dual to the volume of an extremal hypersurface which has a fixed boundary as a time slice on the boundary of AdS space \cite{ref6, ref7, ref8, ref9}. The other proposal is the complexity=action (CA) which corresponds to the gravitational action on the Wheeler-DeWitt (WDW) patch \cite{ref10, ref11}. Both the volume in CV conjecture and WDW patch in CA have the same time slice as boundary and are extended to the interior of black hole. So people believe that the study of holographic complexity may shed lights on the black hole physics.

On the other hand, in the field theory side we need more efforts to understand the complexity better. In a quantum theory, it can be roughly defined as the number of operations needed to transform a reference state $|\psi_{R}\rangle$ to a target state $|\psi_{T}\rangle$. These operators can be called quantum gates and more needed gates means more complicated target state. This brings some kind of distance notion in the Hilbert space. When come to the quantum field theory, there is a very nice suggestion for distance by Nielsen et al \cite{ref18, ref19, ref20}. To be concrete, consider $``s"$ as an affine parameter in a unitary operator $U(s)$ such that $U(s)|\psi_R\rangle$ approches $|\psi_T\rangle$ as $s$ goes from 0 to 1. The unitary operator $U(s)$ can be written in terms of quantum gates $M_I$'s as 
\begin{equation}
	U(s)=\mathcal{P} \exp{ \bigg[ -\int_{0}^{s}ds Y^{I}(s) M_{I} \bigg]},
\end{equation}
where $\mathcal{P}$ denotes the path order of the parameter $s$ and functions $Y^I(s)$'s charactrize the quantum circuit. These functions are chosen such that we reach to a minimal circuit depth or cost function which is defined by
\begin{equation}\label{cost}
	D[U]:=\int_{0}^{1}ds \sqrt{\sum_{I}|Y^I (s)|^{2}}.
\end{equation}
Then one can define the complexity $\mathcal{C}$ as
\begin{equation}
	\mathcal{C}:=\min_{\{Y^{I}\}} D[U].
\end{equation}
in which minimum means varying $Y^I$'s functions to minimize the cost function $D[U]$. 

In this regards, several studies were done including free scalars \cite{ref12, ref13},  free fermions \cite{ref14, ref15} and interacting scalars \cite{ref16} (see also \cite{ref12} which modifies \eqref{cost} by removing the square root sign). In this paper, we are going to consider $U(1)$ vector gauge fields. We apply the covariance matrix approach to evaluate the complexity in field theory. We regard the transformation between the reference and target states as the trajectories in the space of states which connect two states. In this approach, the trajectories can be parametrized as $U(s)=e^{sA}$ which is satisfying the boundary conditions $U(s=0)=1$ and $U(s=1)=e^{A}$. Amongst infinite family of trajectories or geodesics, only the shortest is desired, because it does achieve the cost function or the complexity in terms of the relative covariance matrix $\Delta$.

The organization of the paper is as follows. In section \ref{sec:setup}, we introduce the model and its Hamiltonian. In section \ref{sec:harmonic}, we discrete the space and find the corresponding simple harmonic oscillator model. Section \ref{sec:covariance} is devoted to the covariance matrix approach and we conclude in section \ref{sec:conclusion}.

\section{The Setup}\label{sec:setup}
To begin, we consider a free vector field of a $U(1)$ gauge theory, in $d+1$ spacetime dimensions. By determining our desired QFT, we have to place the theory on a lattice in order to regulate it. In which case, we face with a reduced system composed of an infinite family of coupled harmonic oscillators \cite{ref12}. To study the circuit complexity, we introduce the target and reference states. Furthermore, we need to identify a set of elementary gates which are the unitary transformations that map the reference state onto the target state.

We begin with the QED Lagrangian, in $d+1$ spacetime dimensions,
\begin{align}
	L=\int d^dx\mathcal{L}=\int d^dx\left[-\frac{1}{4} F^{\mu\nu} F_{\mu\nu}+J^\mu A_\mu\right]
\end{align}
For free U(1) gauge theory in vacuum, ($J^\mu=0$),
\begin{align}
	\mathcal{L}=-\frac{1}{4} F^{\mu\nu} F_{\mu\nu}=-\frac{1}{2} \partial^\mu A^\nu \partial_\mu A_\nu+\frac{1}{2} \partial^\mu A^\nu \partial_\nu A_\mu
\end{align}
The time derivative of $A^0$ does not appear in the above Lagrangian. This means that the field includes neither canonically conjugate momentum nor dynamics. The reason is that not all degrees of freedom $A^\mu$, namely all components of field A, are physical since they are subjected to the gauge transformations. Gauge fixing, which eliminates the gauge freedom, is a solution to this problem. 
We choose Coulomb gauge, $\vec{\nabla}\cdot \vec{A}(x)=0$ accompanied by the temporal gauge $A^0=0$. The Lagrangian then becomes as follows:
\begin{align}
	\mathcal{L}=\frac{1}{2} \dot{A}_i \dot{A}^i-\frac{1}{2} \partial^i A^j \partial_i A_j
\end{align}
We need to carry out the Hamiltonian formalism now. We first compute the canonically conjugate momentum to $A_i$,
\begin{align}
	\Pi_i=\partial L/(\partial \dot{A}^i )=\dot{A}_i
\end{align}
The Hamiltonian density is then
\begin{align}
	\mathcal{H}&=\Pi^i \dot{A}_i-\mathcal{L}   \\
&=\frac{1}{2} \Pi^i \Pi_i+\frac{1}{2} \partial^i A^j \partial_i A_j
\end{align}
This is the Hamiltonian density for free photons. Therefore, in $d+1$ spacetime dimensions the Hamiltonian becomes:
\begin{align}
	H=\int d^d x \mathcal{H}=
	\frac{1}{2} \int d^d x \Big[\Pi^i \Pi_i+\partial^i A^j \partial_i A_j \Big] 
\end{align}
The Hamiltonian does not contain the mass term, because field $A^\mu (x)$ (photon) has no mass. Therefore, it is difficult to continue modeling with coupled harmonic oscillators. To solve this problem, we assume that the photon or field $A^\mu$ has very little mass m and finally we take the $m\to 0$ limit. Therefore, the Hamiltonian would be contained a mass term of the photon, as
\begin{align}
	H=\frac{1}{2} \int d^d x (\Pi^i \Pi_i+\partial^i A^j \partial_i A_j+m^2 A^i A_i ) \,,  \qquad     m\to 0
\end{align}
In the first step of aforementioned process, we place the theory on a square lattice with lattice spacing $\delta$ for the sake of regulation. It follows then
\begin{align}\label{eq13}
	H=\frac{1}{2} \delta^d \sum_{\vec{x},i}\left[\Pi_i (\vec{x})^2+\frac{1}{\delta^2}  (A_i (\vec{x}+\delta \hat{x}_i)-A_i (\vec{x}))^2+m^2 A_i (\vec{x})^2 \right]
\end{align}
where $\delta \hat{x}_i$ are unit vectors in the direction of the spatial dimensions of the lattice. Therefore, our system naturally changes to a quantum mechanical problem with an infinite family of coupled harmonic oscillators. By expanding field $\vec{A}(x)$ in terms of the creation and annihilation operators $(a,a^\dagger)$,
\begin{align}\label{eq14}
	\vec{A}(x)=\sum_\lambda\sum_k\left[\varepsilon_{\lambda,k}^* a_{\lambda,k} e^{i\vec{k}\cdot\vec{x}}+\varepsilon_{\lambda,k} a_{\lambda,k}^\dagger e^{i\vec{k}\cdot\vec{x}} \right]
\end{align}
in which $\varepsilon_{\lambda,k}$ is the polarization vector and $\lambda$ is the number of degrees of freedom of the photon. The Coulomb gauge condition implies that the polarization vectors $\varepsilon_{\lambda,k}$ are orthogonal to the wave vector $\vec{k}$, so  $\lambda$  is 1 to $d-1$, $\lambda=\{1,2,\dots,d-1\}$, in d spatial dimensions. Applying the orthogonality of polarization vectors $\varepsilon_{\lambda,k}$ to the wave vector $\vec{k}$, we can rewrite the above Hamiltonian as follows:
\begin{align}\label{eq15}
	H=\sum_\lambda\sum_k\omega a_{\lambda,k}^\dagger a_{\lambda,k} +\text{const.}
\end{align}
where $\omega$ is the energy.

\section{Modeling to the simple harmonic oscillator Hamiltonian}\label{sec:harmonic}
Following \cite{ref12}, the Hamiltonian of N coupled harmonic oscillators on a one-dimensional circular lattice of length $L=N\delta$ in the basis of the position is in the following form
\begin{align}\label{eq16}
	H=\frac{1}{2} \sum_a\Big[p_a^2+\omega^2 q_a^2+\Omega^2 (q_a-q_{a+1} )^2 \Big]     
\end{align}
with periodic boundary conditions $q_{a+N}=q_a$. We have set masses $M_a=1$ for simplicity. Also, we had better consider the frequencies to be associated with the field theory parameters by $\omega=m$ and $\Omega=1/\delta$, as in eq. \eqref{eq13}. Of course, we could simply rewrite the Hamiltonian in terms of the normal modes,
\begin{align}\label{eq17}
	H=\frac{1}{2} \sum_k(|\tilde{p}_k |^2+\tilde{\omega}_k^2 |\tilde{q}_k |^2 )     
\end{align}
where one would achieve the normal-mode basis by a discrete Fourier transform
\begin{align}
	\tilde{q}_k\equiv \frac{1}{\sqrt{N}} \sum_a \exp\Big(-\frac{2\pi ik}{N} a\Big) q_a
\end{align}
where $k=\{1,\cdots,N\}$, and we note that $\tilde{q}_k^\dagger=\tilde{q}_{N-k}=\tilde{q}_{-k}$. Using the above Fourier transform, we can easily derive the normal-mode frequencies $\tilde{\omega}_k$ in the Hamiltonian \eqref{eq17} in terms of $\omega$ and $\Omega$ as follows,
\begin{align}\label{eq19}
	\tilde{\omega}_k^2=\omega^2+4\Omega^2\sin^2\Big(\frac{\pi k}{N}\Big)     
\end{align}
As we know, the Hamiltonian of Eq. \eqref{eq15} is the familiar form of the Hamiltonian of the simple harmonic oscillators; to see, we can write the creation and annihilation operators as follows:
\begin{align}
	a_k^\dagger=\frac{1}{\sqrt{2\omega}} (ip_k+\omega_k q_k ) \,; \qquad       a_k=\frac{1}{\sqrt{2\omega}} (-ip_k+\omega_k q_k )
\end{align}
Therefore, if the Hamiltonian is written in the form of Eq. \eqref{eq15}, it will be in normal mode and changes the problem to a system of N decoupled harmonic oscillators. Thus, the ground-state wave function can be easily written as the product of the ground-state wave functions of individual oscillators,
\begin{align}
	\psi_0 (\tilde{q}_1,\tilde{q}_2,\tilde{q}_3,\cdots)=\prod_{k=1}\Big(\frac{\tilde{\omega}_k}{\pi} \Big)^{\frac{1}{4}}  \exp\Big[-\frac{1}{2} \tilde{\omega}_k |\tilde{q}_k |^2 \Big]
\end{align}
We note that this Gaussian wave function establishes a suitable family of target states in our complexity computations. Since our lattice is $d$ dimensional, there are $N^d$ lattice sites, $N$ sites in each dimension. Photons can place on these sites, and every photon has $d-1$ degrees of freedom. So, the total number of degrees of freedom will be $(d-1)N^d$. Associated with any degree of freedom, there exists an oscillator with $(q_i,p_i)$. Therefore, the range of label $k$ in $\psi_0$ is $k=\{1,2,\cdots,(d-1)N^d \}$. Then, our target state would be as follows:
\begin{align}
	\psi_T=\prod_{k=1}^{(d-1)N^d}\Big(\frac{\tilde{\omega}_k}{\pi} \Big)^{\frac{1}{4}}  \exp\Big[-\frac{1}{2} \tilde{\omega}_k |\tilde{q}_k |^2 \Big]     
\end{align}
Since the ground state wavefunction of the oscillators we have chosen as the target state is a Gaussian function, we choose the reference state in such a way that it is also Gaussian and can be prepared with simple gates. In the next subsection, we discuss about choosing the reference state in more detail. Hence, the desired unitary transformations are those that maps Gaussian states to Gaussian states and is known as Bogoliubov transformation. In another word, the Bogoliubov transformation preserves the form of states.

\section{Covariance matrix approach}\label{sec:covariance}
In the following, we apply the approach of \cite{ref15} and use the covariance matrix to parametrize the Gaussian states. In particular, the two-point function of Gaussian states, or actually any state, can be expressed as \cite{ref15}
\begin{align}\label{eq23}
	\langle\psi|\xi^a \xi^b|\psi\rangle= \frac{1}{2} (G^{ab}+i\Omega^{ab}) 
\end{align}
where $\xi^a=\{q_1,\cdots,q_{(d-1)N^d },p_1,\cdots,p_{(d-1)N^d} \}$ is the coordinates of configuration space and describes $(d-1)N^d$ degrees of freedom. In the right-hand side of Eq. \eqref{eq23}, $G^{ab}$ is symmetric and $\Omega^{ab}$  antisymmetric part that can be achieved as
\begin{align}
	G^{ab}&=\langle\psi|\xi^a \xi^b+\xi^b \xi^a|\psi\rangle    \\
\Omega^{ab}&=\langle\psi|\xi^a \xi^b-\xi^b \xi^a|\psi\rangle
\end{align}
For simplicity, we first consider one degree of freedom $\xi^a=\{q,p\}$, i.e., one oscillator. That is, we start from the following Gaussian state
\begin{align}
	\psi(\tilde{q})=\Big(\frac{\tilde{\omega}}{\pi}\Big)^{\frac{1}{4}} \exp\Big[-\frac{1}{2} \tilde{\omega}|\tilde{q}|^2 \Big]
\end{align}
Hence, the entries of $G^{ab}$ matrix can be obtained as
\begin{align}
	G^{11}&=2\langle\psi|qq|\psi\rangle=2\langle\psi|q^2|\psi\rangle=\frac{1}{\tilde{\omega}}  \\
G^{22}&=2\langle\psi|pp|\psi\rangle=2\langle\psi|p^2|\psi\rangle=\tilde{\omega} \\
G^{12}&=2\langle\psi|qp+pq|\psi\rangle=0   \\
G^{21}&=2\langle\psi|pq+qp|\psi\rangle=0
\end{align}
So, we can write
\begin{align}\label{eq31}
	G=\begin{bmatrix}
		1/\tilde{\omega}  & 0 \\ 0 & \tilde{\omega}
	\end{bmatrix}
\end{align}
The same procedure is applied to derive the  $\Omega^{ab}$ matrix
\begin{align}\label{eq32}
	\Omega=\begin{bmatrix}
		0  & 1 \\ -1 & 0
	\end{bmatrix}
\end{align}
Eqs. \eqref{eq31} and \eqref{eq32} reveal that the $\Omega^{ab}$ matrix is trivial and only the $G^{ab}$ matrix, the symmetric part of the covariance matrix, completely characterizes the Gaussian state. In other words, it contains all physical features of Gaussian states.
Therefore, the covariance matrix associated to our target state become
\begin{align}
G_T=\begin{bmatrix}
		\frac{1}{\tilde{\omega}_1}  &  &  &  &  &  \\ 
		     &  \ddots            &  &  &  0  &  \\
		  &  &   \frac{1}{\tilde{\omega}_{(d-1)N^d}}  &   &   &   \\
		  &  &  & \tilde{\omega}_1  &   &   \\
		  &0 &  &  &  \ddots &   \\
		  &  &  &  &  &  \tilde{\omega}_{(d-1)N^d}
	\end{bmatrix}
\end{align}
which is a $(d-1)N^d\times(d-1)N^d$ dimensional matrix.

According to \cite{ref17} and \cite{ref18}, the power of covariance matrix formulation lies in the fact that we can study trajectories in the state space entirely in terms of $G^{ab}$; provided that the states remain Gaussian. Being limited to these states means that we focus on a specific subgroup of unitary transformations that maps Gaussian states among themselves, i.e., the Bogoliubov transformations. Therefore, the desired transformation that construct our quantum circuit is as follows
\begin{align}
	|G_T\rangle=\hat{U}|G_R\rangle
\end{align}
in which $|G_R\rangle$ and $|G_T\rangle$ are the reference and target states, respectively. In language of the covariance matrix, we have
\begin{align}
	G_T=UG_R U^T
\end{align}
where, as we know $G_R$ and $G_T$ are the covariance matrices associated with their reference and target states. One can check that the matrix $U$ is not unitary and belongs to the symplectic group $Sp\big(2(d-1)N^d,R\big)$ \cite{ref17}. These trajectories in the state space, which are the geodesics between reference and target states, can be parametrized by the matrix $U$ as $U(s)=e^{sA}$. Therefore, our quantum circuits of interest are these trajectories satisfying the boundary conditions $U(s=0)=1$ and $U(s=1)=e^A$. Amongst infinite family of trajectories or geodesics which connect $G_R$ to $G_T$, only the shortest geodesic is desired, because it does achieve the cost function or the complexity. Hence, the shortest geodesic is given by
\begin{align}\label{eq36}
	\gamma:[0,1]\to Sp\big(2(d-1) N^d,R\big):  s\to e^{sA}  \quad   \text{with}\quad    A=\frac{1}{2}  \log \Delta
\end{align}
where $\Delta$ is the relative covariance matrix which is defined as
\begin{align}\label{eq37}
	\Delta^a_b=(G_T )^{ac} (g_R )_{cb}
\end{align}
where $g$ is the inverse of G, such that $G^{ac} g_{cb}=\delta^a_{\;b}$.
Having identified the shortest geodesic which connects $G_R$ to $G_T$, the circuit complexity is defined as in the following 
\begin{align}\label{eq38}
	\mathcal{C}_2 (G_R,G_T )=||A||=\frac{1}{2} \sqrt{Tr [(\log\Delta )^2 ]}
\end{align}
To calculate the matrix $\Delta$, we choose a basis such that the matrix representation of $G_R$ becomes the identity, i.e., $G_R=1$. Thus, from Eq. \eqref{eq37} we have
\begin{align}\label{eq39}
	\Delta=G_T
\end{align}
Therefore, from Eqs. \eqref{eq36} and \eqref{eq39} the matrix $A$ derive as follows
\begin{align}
	A=\frac{1}{2} \log\Delta=\frac{1}{2}\begin{bmatrix}
		-\log{\tilde{\omega}_1}  &  &  &  &  &  \\ 
		&  \ddots            &  &  &  0  &  \\
		&  &   -\log{\tilde{\omega}_{(d-1)N^d}}  &   &   &   \\
		&  &  & \log\tilde{\omega}_1  &   &   \\
		& 0 &  &  &  \ddots &   \\
		&  &  &  &  &  \log\tilde{\omega}_{(d-1)N^d}
	\end{bmatrix}
\end{align}
Therefore, Eq. \eqref{eq38} yield the complexity as
\begin{align}
	\mathcal{C}_2 (G_R,G_T )=\frac{1}{2} \sqrt{\sum_{\{k_i \}=1}^{(d-1)N}\Big(\log\tilde{\omega}_{\vec{k}}\Big)^2 }        
\end{align}
where $\vec{k}=\{k_1,k_2,\cdots,k_d\}$. Now, in the limit $m\to 0$, Eq. \eqref{eq19} becomes
\begin{align}
	\tilde{\omega}_k^2=4\Omega^2 \sum_{i=1}^d\sin^2\Big(\frac{\pi k_i}{N}\Big)        
\end{align}
The experience with quantum field theory implies that the above equation would be dominated by the UV modes, i.e., the modes $\tilde{\omega}_k \sim 1/\delta$. Hence, we employ the approximation to determine the leading contribution to our circuit complexity as follows
\begin{align}\label{result}
	\mathcal{C}_2 (G_R,G_T )\approx\frac{\sqrt{d-1} N^{d/2}}{2}
	\log\frac{1}{\delta} \sim \Big(\frac{(d-1)V}{\delta ^d} \Big)^{\frac{1}{2}}
\end{align}
where we have used $N^d=V⁄\delta^d$  to re-write the leading power of $N$ in terms of $V⁄\delta^d$. This result is equivalent to the complexity of $d-1$ free scalar fields. The $(d-1)$ factor corresponds to number of degrees of freedom of our gauge vector field. In other words, the $U(1)$ gauge field complexity is equivalent to the $(d-1)$ free scalar fields one.

\section{Conclusion}\label{sec:conclusion}
In this paper, we considered a $U(1)$ gauge field and derive its complexity in the geometric approach of Nielsen. We restrict the problem to Gaussian states and use the Bogoliubov transformations to construct the quantum gates, then we found the complexity as a geodesic distance in the corresponding group manifold of unitary transformations. 

It is well-known that introducing a Hamiltonian for gauge fields requires removing the excessive degrees of freedom due to the gauge invariance. It can be done by gauge fixing. We select the Coulomb gauge and finally found the complexity as \eqref{result} which is equivalent to $(d-1)$ number of free scalar fields \cite{ref12,ref13} somehow indicating that the complexity is related to the number of degrees of freedom. 

Remind that in our model we consider a tiny mass $m$ for the gauge field to avoid unwanted divergences in the harmonic oscillator modeling and in the end of the day we sent $m\to 0$. It is also possible to keep this mass nonzero to find the complexity of a massive Proca vector field. In this case, we will find the complexity as 
\begin{align}\label{proca}
	\mathcal{C}_{Proca}\approx\frac{\sqrt{d} N^{d/2}}{2}
	\log\frac{1}{\delta} \sim \Big(\frac{d\;V}{\delta ^d} \Big)^{\frac{1}{2}}
\end{align}
which differs from \eqref{result} by replacing the factor $(d-1)$ by $d$. This is because of the fact that massive vector field has an extra degree of freedom which is the longitudinal polarization.

It is worth mentioning that the result \eqref{result} is not necessarily the same as the holographic calculations as this is the case in free scalar theories. Following \cite{ref12}, the resolution is to modify the cost function by removing the square root sign from  \eqref{cost}.

There are some interesting issues worth to explore. Firstly, is our complexity depends on gauge fixing? In our model we considered the Coulomb gauge. So if the answer to this question is positive then one may ask what the complexity is in other gauges. Otherwise, if the complexity is gauge invariant one may ask: Is it possible to find a gauge covariant method to compute the complexity?  The second interesting problem is calculation of the complexity for non-Abelian gauge theories and would be a nice extension of our work. Of course, due to the interaction between non-Abelian fields, it will be a bridge to the exploration of circuit complexity of the interaction field theories.

{\bf Note added:} While this paper is under review, we received \cite{Meng:2021wmz} which has some overlaps with our results.



\end{document}